\documentclass[a4paper,12pt]{article}

\usepackage{amsmath,amssymb,latexsym}
\usepackage[dvips]{graphicx}

\newcommand{\ftilde}{\tilde{\phi}}
\newcommand{\kp}{\kappa}

\newcommand{\xhat}{\hat{x}}

\newcommand{\Phat}{\hat{P}}

\newcommand{\kbar}{\bar{k}}
\newcommand{\Pbar}{\bar{P}}
\newcommand{\stilde}{\tilde{\sigma}}
\newcommand{\calJ}{\mathcal{J}}
\newcommand{\calP}{\mathcal{P}}

\newcommand{\tar}{\triangleright}

\newcommand{\Acal}{\mathcal{A}}

\begin{document}
\renewcommand{\thefootnote}{\fnsymbol{footnote}}
\begin{titlepage}
\hfill
{\hfill \begin{flushright} 
YITP-09-09
\end{flushright}
  }

\vspace*{10mm}

\begin{center}
{\LARGE {\LARGE  
The Cutkosky rule of three dimensional noncommutative field theory
in Lie algebraic noncommutative spacetime}}
\vspace*{18mm}

 {\large Yuya Sasai\footnote{ e-mail: sasai@yukawa.kyoto-u.ac.jp}
 and Naoki Sasakura\footnote{e-mail: sasakura@yukawa.kyoto-u.ac.jp}}

\vspace*{13mm}
{\large {\it 
Yukawa Institute for Theoretical Physics, Kyoto University, \\ 
 Kyoto 606-8502, Japan 
}} \\

\end{center}

\vspace*{15mm}

\begin{abstract}
We investigate the unitarity of three dimensional noncommutative scalar field theory in the Lie algebraic noncommutative spacetime $[\xhat^i, \xhat^j]=2i\kappa \epsilon^{ijk}\xhat_k$. This noncommutative field theory possesses an $SL(2,R)/Z_2$ group momentum space, which leads to a Hopf algebraic translational symmetry. We check the Cutkosky rule of the one-loop self-energy diagrams in the noncommutative $\phi^3$ theory when we include a braiding, which is necessary for the noncommutative field theory to possess the Hopf algebraic translational symmetry at quantum level. Then, we find that the Cutkosky rule is satisfied if the mass is less than $1/\sqrt{2}\kappa$.
\end{abstract}

\end{titlepage}

\newpage
\renewcommand{\thefootnote}{\arabic{footnote}}
\setcounter{footnote}{0}

\section{Introduction}
Noncommutative field theories \cite{Snyder:1946qz,Yang:1947ud,Connes:1990qp,Doplicher:1994tu} are interesting subjects which possess the connections with Planck scale physics, such as string theory and quantum gravity. 
The most well-studied are noncommutative field theories in the Moyal spacetime, whose coordinate commutation relation is given by 
$[\xhat^{\mu},\xhat^{\nu}]=i\theta^{\mu\nu}$ with an antisymmetric constant $\theta^{\mu\nu}$. 
Such field theories are known to appear as effective field theories of open string theory with a constant background $B_{\mu\nu}$ field \cite{Connes:1997cr,Seiberg:1999vs}. Various aspects have been extensively analyzed not only as the simplest field theories 
in quantum spacetime but also as toy models of string theory \cite{Douglas:2001ba,Szabo:2001kg}. 

Recently, it has been pointed out that the Moyal spacetime is invariant under the twisted Poincar\'e transformation \cite{Chaichian:2004za,Wess:2003da,Koch:2004ud}, which has a Hopf algebraic structure; the Leibnitz rule of the symmetry algebra is deformed \cite{Majid:1996kd,Klimyk:1997eb}. To implement the twisted Poincar\'e invariance in the noncommutative field theories at quantum level, it has been found that one has to impose a nontrivial statistics on fields, which is called braiding \cite{Oeckl:2000eg,Balachandran:2005eb}. In fact, we can demonstrate that in general setting, for correlation functions to possess a Hopf algebraic symmetry at quantum level, we have to include a braiding \cite{Sasai:2007me}.

Since the Moyal phase is canceled by the braiding \cite{Balachandran:2005eb}, the nonplanar amplitudes, which usually violate the unitarity when the timelike noncommutativity does not vanish \cite{Gomis:2000zz,Aharony:2000gz,AlvarezGaume:2001ka}, trivially satisfy the Cutkosky rule \cite{Cutkosky:1960sp,Peskin:1995ev} if we include the braiding.

In this paper, we study three dimensional noncommutative scalar field theory in the Lie algebraic noncommutative spacetime $[\xhat^i, \xhat^j]=2i\kappa \epsilon^{ijk}\xhat_k~(i,j,k=0,1,2)$ \cite{Imai:2000kq,Freidel:2005bb}. This noncommutative field theory is also physically interesting because the Euclidean version of the theory is known to appear as the effective field theory of three dimensional quantum gravity theory (Ponzano-Regge model \cite{Ponzano:1968}) which couples with spinless massive particles \cite{Freidel:2005bb}. Since massive particles coupled with three dimensional Einstein gravity are understood as conical singularities in three dimensions \cite{Deser:1983tn}, this noncommutative field theory is expected to describe the dynamics of such conical singularities.

We investigate the unitarity of the three dimensional noncommutative scalar field theory in the Lie algebraic noncommutative spacetime. This noncommutative field theory also possesses a Hopf algebraic translational symmetry \cite{Freidel:2005bb,Sasai:2007me,Sasai:2007mc}, since the momentum space has an $SL(2,R)/Z_2$ group structure, which has been shown 
based on the assumptions of commutative momentum operators and Lorentz invariance \cite{Imai:2000kq}. As mentioned above, for the Hopf algebraic translational symmetry to hold in the noncommutative field theory at quantum level, we have to introduce braiding among fields \cite{Freidel:2005bb,Sasai:2007me,Sasai:2007mc}. With the braiding, the nonplanar amplitudes become the same as the corresponding planar amplitudes if they exist. But unlike the Moyal case, even the planar amplitudes are nontrivial because of the nontrivial momentum space. Thus, it is a non-trivial issue whether the Cutkosky rules for various planar as well as non-planar amplitudes hold in the Lie-algebraic noncommutative field theory,
even when the braiding is introduced.

This paper is organized as follows. In section \ref{sec:ncft}, we review the three dimensional noncommutative scalar field theory in the Lie algebraic noncommutative spacetime. In section \ref{subsec:comrel}, we explain why the noncommutative field theory possesses the $SL(2,R)/Z_2$ group momentum space. There are two approaches to construct the noncommutative field theory. In section \ref{subsec:star}, we review the star product formalism. In section \ref{subsec:operator}, we explain the operator formalism. In section \ref{subsec:hopftrans}, we explain the Hopf algebraic translaitonal symmetry in the noncommutative field theory. In section \ref{sec:cut}, we investigate the unitarity of the noncommutative field theory in the Lie algebraic noncommutative spacetime. In section \ref{subsec:amplitude}, we calculate the one-loop self-energy diagrams of the noncommutative scalar $\phi^3$ theory. In section \ref{subsec:unitarity}, we check whether the Cutkosky rule is satisfied at the one-loop self-energy diagrams when we consider the braiding and show that the Cutkosky rule holds when the mass $M$ is less than $1/\sqrt{2}\kappa$. The final section is devoted to a summary and a comment.

\section{Three dimensional noncommutative  field theory in the Lie algebraic noncommutative spacetime} \label{sec:ncft}
In this section, we review a three dimensional noncommutative scalar field theory in the Lie algebraic noncommutative spacetime whose commutation relation is given by
\begin{equation}
[\xhat^i,\xhat^j]=2i\kappa \epsilon^{ijk} \xhat_k, \label{eq:comx}
\end{equation}
where $i,j,k=0,1,2$ \cite{Sasakura:2000vc,Madore:2000en},\footnote{The signatures of the metric and the totally antisymmetric tensor are the following:
\begin{align}
\eta^{ij}&=(-1,1,1), \notag \\
\epsilon^{012}&=1. \notag
\end{align}} 
following the constructions of 
\cite{Imai:2000kq,Freidel:2005bb}.

\subsection{Commutation relations and the momentum space} \label{subsec:comrel}
At first, we assume the following things:
\begin{itemize}
	\item The momentum operators are commutative: $[\Phat^i, \Phat^j]=0$.
	\item The three dimensional Lorentz invariance.
	\item The Jacobi identity.
	\item The commutation relations of $\xhat^i$ and $\Phat^i$ satisfy the ordinary canonical commutation relation in $\kappa\to 0$ limit.
\end{itemize}
Then, we can uniquely determine the commutation relations of $\xhat^i$ and $\Phat^i$ as 
\begin{equation}
[\Phat^i,\xhat^j]=-i\eta^{ij}\sqrt{1+\kappa^2\Phat^i\Phat_i}+i\kappa\epsilon^{ijk}\Phat_k, \label{eq:comp}
\end{equation}
up to the redefinition $P^i\to f(\kappa^2P^jP_j)P^i$, where $f$ is an arbitrary function \cite{Sasakura:2000vc}.
By identifying $\xhat^i$ and $\Phat^i$ with the $ISO(2,2)$ Lie algebra as\footnote{The remaining three independent operators 
\[\hat{M}_i\equiv -\frac{1}{2}\epsilon_{i}{}^{jk}\calJ_{jk}\]
are understood as the $SO(2,1)$ Lorentz generators of the noncommutative spacetime.}
\begin{align}
&\xhat_i=\kappa(\calJ_{-1,i}-\frac{1}{2}\epsilon_{i}{}^{jk}\calJ_{jk}), \label{eq:xhat} \\
&\Phat_i=\calP_{\mu=i}, 
\end{align}
and imposing the constraint
\begin{equation}
\calP_{-1}=\frac{1}{\kappa}\sqrt{1+\kappa^2\calP^i\calP_i}, \label{eq:operatorconstraint}
\end{equation}
we can show that the commutation relations (\ref{eq:comx}) and (\ref{eq:comp}) can be derived from the $ISO(2,2)$ Lie algebra \cite{Imai:2000kq}. Here, the commutation relations of $ISO(2,2)$ Lie algebra are\footnote{The Greek indices run through $-1$ to $2$ and $\eta^{\mu\nu}=(-1,-1,1,1)$.} 
\begin{align}
[\calJ_{\mu\nu},\calJ_{\rho\sigma}]&=-i(\eta_{\mu\rho}\calJ_{\nu\sigma}-\eta_{\mu\sigma}\calJ_{\nu\rho}-\eta_{\nu\rho}\calJ_{\mu\sigma}+\eta_{\nu\sigma}\calJ_{\mu\rho}), \label{eq:comjj} \\
[\calJ_{\mu\nu},\calP_{\rho}]&=-i(\eta_{\mu\rho}\calP_{\nu}-\eta_{\nu\rho}\calP_{\mu}), \label{eq:comjp} \\
[\calP_{\mu},\calP_{\nu}]&=0.
\end{align}

\begin{figure}
\begin{center}
\includegraphics[scale=0.55]{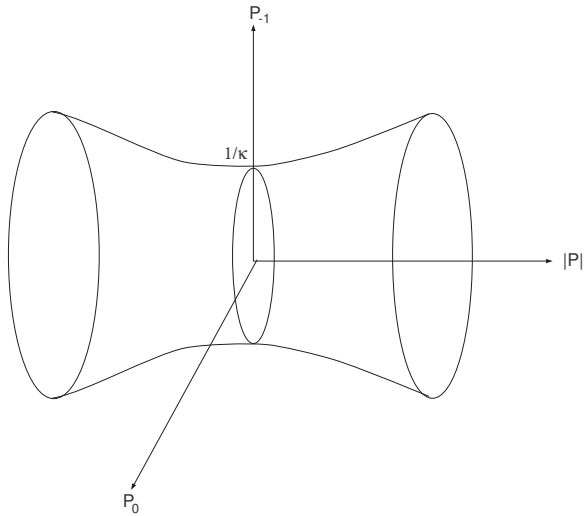}
\end{center}
\caption{The momentum space, which is the same as $AdS_3$ space with a radius $1/\kappa$. Here, $|P|\equiv \sqrt{P_1^2+P_2^2}$. }
\label{fig:ads3}
\end{figure}

Since the momentum operators are commutative and follow the constraint (\ref{eq:operatorconstraint}), a representation space of the Lie algebra can be given by functions of momenta on the following hyperboloid,
\begin{equation}
P^{\mu}P_{\mu}=-\frac{1}{\kappa^2}, \label{eq:hyperboloid}
\end{equation}
depicted as in Figure \ref{fig:ads3}. 

Then, we can identify the momentum space with an $SL(2,R)$ group manifold as follows:\footnote{The $\tilde{\sigma}^i$s are defined by \[\tilde{\sigma}^0=
\sigma^2, ~\tilde{\sigma}^1=i\sigma^3,~ \tilde{\sigma}^2=i\sigma^1,\] with Pauli 
matrices
\[
\sigma^1=
\begin{pmatrix}
0 & ~1 \\
1 & ~0 \\
\end{pmatrix},~
\sigma^2=
\begin{pmatrix}
0 & ~-i \\
i & ~0 \\ 
\end{pmatrix},~
\sigma^3=
\begin{pmatrix}
1 & 0 \\
0 & ~-1 \\
\end{pmatrix}.
\]
 These matrices satisfy
\[
 \stilde^i\stilde^j=-\eta^{ij}+i\epsilon^{ijk}\stilde_k.
\]
}
\begin{equation}
g=P_{-1}(g)+i\kappa P_i(g)\tilde{\sigma}^i, ~~~~\det g=1, \label{eq:gdef1}
\end{equation}
because the determinant condition of $g$ is equivalent to
\begin{equation}
P_{-1}(g)^2-\kappa^2P^i(g)P_i(g)=1, \label{eq:defofp-1}
\end{equation}
which is the same as (\ref{eq:hyperboloid}) with the identification of $P_{-1}(g)= \kappa P_{-1}$.

$P_{-1}(g)$ in (\ref{eq:defofp-1}) has two-fold degeneracy for each $P_i(g)$. To delete this physically unwanted degeneracy, we impose an identification condition on a field, which we will see in the next section.

The expression (\ref{eq:defofp-1}) implies that the mass
$M^2=-P^i(g)P_i(g)$ has an upper bound given by
\begin{equation}
M^2\leq \frac{1}{\kappa^2}.
\end{equation}

\subsection{The star product formalism} \label{subsec:star}
Next, we review the star product formalism of the noncommutative scalar field theory in the Lie algebraic noncommutative spacetime, developed in \cite{Freidel:2005bb}. We take the momentum space $g$ as $SL(2,R)$.\footnote{If we take $g$ as $SL(2,R)/Z_2$, we can not construct the well-defined star product \cite{Joung:2008mr}.} 

We define a scalar field $\phi(x)$ through Fourier transformation of $\ftilde(g)$ as follows:
\begin{equation}
\phi(x)=\int dg \tilde{\phi}(g) e^{iP(g)\cdot x}, \label{eq:phistarf}
\end{equation}
where  $dg$ is the Haar measure of $SL(2,R)$ and $P(g)\cdot x\equiv P(g)_i x^i$. 

The star product is defined as\footnote{We can reproduce the commutation relation  (\ref{eq:comx}) 
by differentiating both hand sides of (\ref{eq:defofstar}) with respect to 
$P_i(g_1)$ and  $P_j(g_2)$ and then taking the limit $P_i(g_1), P_j(g_2)\to 0$.}
\begin{equation}
e^{iP(g_1)\cdot x}\star e^{iP(g_2)\cdot x}=e^{iP(g_1g_2)\cdot x}, \label{eq:defofstar}
\end{equation}
where
\begin{align}
P_i(g_1g_2)&=P_i(g_1)P_{-1}(g_2)+P_{-1}(g_1)P_i(g_2)-\kp \epsilon_{i}{}^{jk}P_j(g_1)P_k(g_2), \label{eq:g1g2i} \\
P_{-1}(g_1g_2)&=P_{-1}(g_1)P_{-1}(g_2)+\kp^2 P^i(g_1)P_i(g_2). \label{eq:g1g2-1}
\end{align}

With these tools, we construct the action of the noncommutative scalar field theory. For example, the action of noncommutative $\phi^3$ theory is given by
\begin{equation}
S=\int d^3x\bigg[-\frac{1}{2}(\partial^i\phi \star \partial_i\phi)(x)-\frac{1}{2}M^2(\phi \star \phi)(x)-\frac{\lambda}{3}(\phi \star \phi \star \phi)(x) \bigg]. \label{eq:action}
\end{equation}
In momentum representation, the action (\ref{eq:action}) becomes
\begin{align}
S=-\frac{1}{2}&\int dg_1dg_2 \ftilde(g_1)(P^2(g_2)+M^2)\ftilde(g_2)(\delta(g_1g_2)+\delta(-g_1g_2)) \notag \\
&-\frac{\lambda}{3}\int dg_1dg_2dg_3\ftilde(g_1)\ftilde(g_2)\ftilde(g_3)(\delta(g_1g_2g_3)+\delta(-g_1g_2g_3)). 
\end{align}
To delete two-fold degeneracy of $P_{-1}(g)$ for each $P_i(g)$, we impose
\begin{equation}
\ftilde(g)=\ftilde(-g). \label{eq:idfphi}
\end{equation}
Then, the action becomes
\begin{align}
S=-&\int dg \ftilde(g^{-1})(P^2(g)+M^2)\ftilde(g) \notag \\
&-\frac{2\lambda}{3}\int dg_1dg_2dg_3\ftilde(g_1)\ftilde(g_2)\ftilde(g_3)\delta(g_1g_2g_3).  \label{eq:actionstar}
\end{align}

In this formalism, if we impose (\ref{eq:idfphi}), we have a complication that $\phi(x)$ defined in (\ref{eq:phistarf}) becomes the same as $\phi(-x)$.  This is not a serious problem since we may become more careful in defining a field in the coordinate $x^i$. In fact, in the next section, we see that such complications are not found in the operator formalism.

\subsection{The operator formalism} \label{subsec:operator}
Next, we review the operator formalism of the noncommutative $\phi^3$ theory, developed in \cite{Imai:2000kq}. An $SL(2,R)$ group element $g$ can be also represented by the exponential of the Pauli matrices $\stilde^i$:
\begin{equation}
g=e^{i\kappa k \cdot \stilde}, \label{eq:gdef2}
\end{equation}
Comparing (\ref{eq:gdef1}) and (\ref{eq:gdef2}), we find the relations between $P_{\mu}$ and $k_i$ as follows:
\begin{align}
P_{-1}&=\cosh (\kappa\sqrt{k^2}), \notag \\
P_i&=k_i\frac{\sinh(\kappa\sqrt{k^2})}{\kappa\sqrt{k^2}}. \label{eq:pkrelation}
\end{align}

A one particle state is given by
\begin{equation}
|g\rangle \equiv e^{ik(g)\cdot \xhat}|0\rangle , \label{eq:oneparticlestate}
\end{equation}
where $|0\rangle $ denotes the zero momentum eigenstate with $P_{-1}=1$. In fact, this state satisfies
\begin{align}
\Phat_ie^{ik(g)\cdot \xhat}|0\rangle &=P_i(g)e^{ik(g)\cdot \xhat}|0\rangle, \\
\Phat_{-1}e^{ik(g)\cdot \xhat}|0\rangle 
&=P_{-1}(g)e^{ik(g)\cdot \xhat}|0\rangle,
\end{align}
where we have used the following formula,
\begin{equation}
\Phat_{\mu} e^{ik(g)\cdot \xhat}=e^{ik(g)\cdot \xhat}T_{\mu}{}^{\nu}(g)\Phat_{\nu}, \label{eq:pecom}
\end{equation}
where 
\begin{equation}
T(g)_{\mu}{}^{\nu}=
\begin{pmatrix}
P_{-1}(g) & ~-P_0(g) & ~P_1(g) & ~P_2(g) \\
P_0(g) & ~P_{-1}(g) & ~-P_2(g) & ~P_1(g) \\
P_1(g) & ~-P_2(g) & ~P_{-1}(g) & ~P_0(g) \\
P_2(g) & ~P_1(g) & ~-P_0(g) & ~P_{-1}(g)
\end{pmatrix}. \label{eq:tmatrix}
\end{equation}
The proof of the formula (\ref{eq:pecom}) is given in the appendix \ref{sec:proofpecom}.
Thus, we find that (\ref{eq:oneparticlestate}) is a state whose momentum is equal to $P_i(g)$ with $P_{-1}(g)$.

We define a scalar field as follows:
\begin{equation}
\phi(\xhat)=\int dg \ftilde(g) e^{ik(g)\cdot \xhat}.
\end{equation}
We impose the condition (\ref{eq:idfphi}) as we have done in the star product formalism. In this formalism, there seems no problem to impose (\ref{eq:idfphi}).

Acting the field on the vacuum $|0 \rangle $, we obtain
\begin{equation}
|\phi \rangle =\int dg \ftilde(g) |g\rangle,
\end{equation}
which is interpreted as a superposition of arbitrary momentum one-particle states.

The product of the plane waves is given by the Baker-Campbell-Haussdorff formula. Since the Baker-Campbell-Haussdorff formula is nothing but the group multiplication, we obtain
\begin{equation}
e^{ik(g_1)\cdot \xhat}e^{ik(g_2)\cdot \xhat}=e^{ik(g_1g_2)\cdot \xhat}.
\end{equation}
Using the above definitions, we can construct the action of the noncommutative $\phi^3$ theory as follows:
\begin{equation}
S=-\frac{1}{2}\langle 0|\phi(\xhat)(\Phat^2+M^2)\phi(\xhat)|0\rangle 
-\frac{\lambda}{3}\langle 0|\phi(\xhat)\phi(\xhat)\phi(\xhat)|0\rangle .
\end{equation}
Using the following formula \cite{Imai:2000kq}:
\begin{equation}
\langle 0|g\rangle =\delta(g),
\end{equation}
the momentum representation of the action is
\begin{align}
S&=-\frac{1}{2}\int dg\ftilde(g^{-1})(P(g)^2+M^2)\ftilde(g) \notag \\
&-\frac{\lambda}{3}\int dg_1dg_2dg_3\ftilde(g_1)\ftilde(g_2)\ftilde(g_3)\delta(g_1g_2g_3), \label{eq:momentumrepaction}
\end{align}
which is essentially the same as (\ref{eq:actionstar}).

\subsection{The Hopf algebraic translational symmetry} \label{subsec:hopftrans}
At first, we briefly review the Hopf algebra and the \textit{action}\footnote{We use italics to distinguish it from the action $S$.} (representation) of Hopf algebra on vector spaces \cite{Majid:1996kd,Klimyk:1997eb}.

A Hopf algebra $\Acal$ is an algebra which is equipped with the following mappings:\begin{align}
&m : \Acal \otimes \Acal \to \Acal~~~\textit{(product)}, \\
&u : \Bbbk \to \Acal~~~\textit{(unit)}, \\
&\Delta : \Acal \to \Acal \otimes \Acal~~~\textit{(coproduct)}, \\
&\epsilon : \Acal \to \Bbbk~~~\textit{(counit)}, \\
&S : \Acal \to \Acal~~~\textit{(antipode)},
\end{align}
which satisfy
\begin{align}
m\circ (m\otimes id)&=m\circ (id\otimes m), ~~~~~~\textit{(associativity)} \\
m\circ (id\otimes u)&=id=m\circ (u\otimes id), \\
(\Delta \otimes id)\circ \Delta&=(id\otimes \Delta)\circ \Delta,~~~~~~\textit{(coassociativity)} \label{eq:coassociative} \\
(id \otimes \epsilon)\circ \Delta&=id=(\epsilon\otimes id)\circ \Delta, \\
m\circ(S\otimes id)\circ\Delta&=u\circ \epsilon=m\circ(id\otimes S)\circ\Delta,
\end{align}
where $\Bbbk$ is a c-number.

An \textit{action} $\alpha_V$ is a map $\alpha_V:\mathcal{A}\otimes V\to V$, where $\mathcal{A}$ is an arbitrary Hopf algebra and $V$ is a vector space. In abbreviated form, we write the \textit{action} of Hopf algebra as $a\triangleright V$, where $a$ is an element of the Hopf algebra. The most important axiom is that an \textit{action} on a tensor product of vector space $V$ and $W$ is defined by
\begin{equation}
a \triangleright (V\otimes W)=\Delta a \triangleright (V\otimes W),
\end{equation}
where $\Delta$ is the coproduct of the Hopf algebra.
If we suppose the coassociativity of a Hopf algebra,
\begin{equation}
(\Delta \otimes id)\circ \Delta(a)=(id \otimes \Delta)\circ \Delta(a), \label{eq:coassociativity}
\end{equation}
the \textit{action} on a tensor product of more than two vector spaces is also uniquely determined.

Next, we explain the Hopf algebraic translational symmetry in the noncommutative field theory.
Let us denote the translational transformation of a field $\ftilde(g)$ as
\begin{equation}
P_{\mu}\tar \ftilde(g)=P_{\mu}(g)\ftilde(g),
\end{equation}
where $P_{\mu}$ are the elements of the (Hopf) algebras of the translation. The \textit{action} of $P_{\mu}$ on the tensor product $\ftilde(g_1)\ftilde(g_2)$ is defined with the coproduct $\Delta$ by
\begin{align}
P_{\mu}\tar \ftilde(g_1)\ftilde(g_2)\equiv \Delta P_{\mu}\tar \ftilde(g_1)\ftilde(g_2).
\end{align}
In the case of the product of three fields, the \textit{action} of $P_{\mu}$ is given by
\begin{align}
P_{\mu}\tar \ftilde(g_1)\ftilde(g_2)\ftilde(g_3)&\equiv (\Delta\otimes id)\circ \Delta P_{\mu}\tar \ftilde(g_1)\ftilde(g_2)\ftilde(g_3) \notag \\
&=(id\otimes \Delta)\circ \Delta P_{\mu}\tar \ftilde(g_1)\ftilde(g_2)\ftilde(g_3).
\end{align}
Similarly, the \textit{action} on arbitrary products of fields is uniquely determined by the coproduct which satisfies the coassociativity (\ref{eq:coassociative}).

In our case, (\ref{eq:g1g2i}) and (\ref{eq:g1g2-1}) determine the coproduct of $P_i$ and $P_{-1}$ as
\begin{align}
\Delta P_i&=P_i\otimes P_{-1}+P_{-1}\otimes P_i-\kp \epsilon_{i}{}^{jk}P_j\otimes P_k, \label{eq:coprop} \\
\Delta P_{-1}&=P_{-1}\otimes P_{-1}+\kp^2 P^i\otimes P_i. \label{eq:coprop-1}
\end{align}
In fact, 
\begin{equation}
\Delta P_{\mu}\tar (\ftilde(g_1)\ftilde(g_2))=P_{\mu}(g_1g_2)\ftilde(g_1)\ftilde(g_2).
\end{equation}
Thus, we find that the coproduct (\ref{eq:coprop}) is different from the usual one,
\begin{equation}
\Delta P_i=P_i\otimes 1+1\otimes P_i, \label{eq:usualcopro}
\end{equation}
which leads to the usual Leibnitz rule. In $\kappa \to 0$ limit, (\ref{eq:coprop}) becomes (\ref{eq:usualcopro}).

Using these coproducts, we can discuss the Hopf algebraic translational symmetry of the noncommutative field theory. For example, let us consider the \textit{action} of $P^i$ on the interaction term of (\ref{eq:momentumrepaction}). Then, it becomes
\begin{align}
&P^i\tar \int dg_1dg_2dg_3 \ftilde(g_1)\ftilde(g_2)\ftilde(g_3)\delta(g_1g_2g_3) \notag \\
&=\int dg_1dg_2dg_3 P^i\tar (\ftilde(g_1)\ftilde(g_2)\ftilde(g_3))\delta(g_1g_2g_3) \notag \\
&=\int dg_1dg_2dg_3 P^i(g_1g_2g_3) (\ftilde(g_1)\ftilde(g_2)\ftilde(g_3))\delta(g_1g_2g_3) \notag \\
&=0.
\end{align}
Thus, the interaction term is invariant under the Hopf algebraic translational symmetry. In the same way, we can show that the total action of the noncommutative field theory is invariant under the Hopf algebraic translational symmetry.

\section{One-loop self-energy amplitudes of the noncommutative field theory and the Cutkosky rule} \label{sec:cut}

\subsection{One-loop self-energy amplitudes of the noncommutative $\phi^3$ theory} \label{subsec:amplitude}
In this section, we review the calculation of the amplitudes in the three dimensional noncommutative scalar field theory in the Lie algebraic noncommutative spacetime \cite{Imai:2000kq}. We can read the Feynman rules from the action (\ref{eq:momentumrepaction}) as follows:\footnote{Strictly speaking, there exists some complications coming from the identification (\ref{eq:idfphi}). For example, the vertex rule should be given by
\[-i\lambda_n (\delta(g_1\cdots g_n)+\delta(-g_1\cdots g_n)).\]
But changing (\ref{eq:feynman rule}) to this vertex rule does not change the essence of the calculations of the amplitudes.}
\begin{align}
\text{propagator}:~~~~~&\frac{-i}{P^2(g)+M^2}, \\
\text{n-vertex}:~~~~~&-i\lambda_n \delta(g_1\cdots g_n). \label{eq:feynman rule}
\end{align}

\begin{figure}
\begin{center}
\includegraphics[scale=1.0]{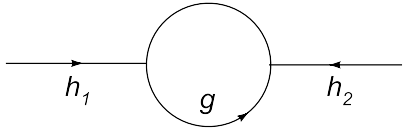}
\end{center}
\caption{The one-loop self-energy planar diagram.}
\label{fig:planar1-loop}
\end{figure}

Using the above rules, we can calculate the loop amplitudes. Let us first show the calculation of the planar one-loop self-energy amplitude in the noncommutative $\phi^3$ theory, which is depicted as in Figure \ref{fig:planar1-loop}.  

\begin{figure}
\begin{center}
\includegraphics[scale=0.6]{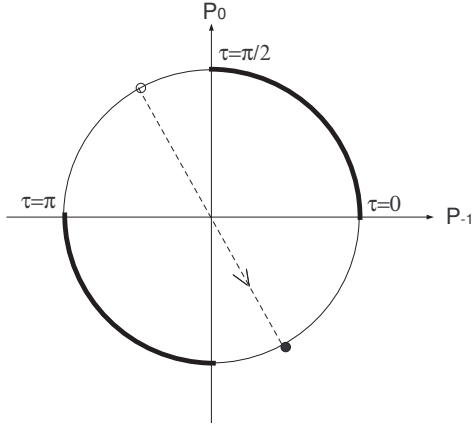}
\end{center}
\caption{The positive energy region is shown by the bold lines. $\tau+\pi$ and $\tau$ are identified because of $\phi(g)\sim \phi(-g)$.}
\label{fig:identify}
\end{figure}

The amplitude of the planar diagram is given by
\begin{equation}
i\Gamma_p^{(2)}=\lambda^2\int dg\delta (h_1h_2)\frac{1}{P^2(g)+M^2}\frac{1}{P^2(h_1^{-1}g)+M^2}. \label{eq:1loopamp}
\end{equation}
For simplicity, we set $\kappa =1$ without loss of generality. Since $SL(2,R)$ group space is equivalent to $AdS_3$ space, we can use the global coordinates,
\begin{equation}
P(g)_{\mu}=(\cosh \rho \cos \tau, \cosh \rho \sin \tau, \sinh \rho \cos \phi, \sinh \rho \sin \phi),
\end{equation}
where $0\leq \rho \leq \infty, ~0\leq \tau \leq 2\pi, ~0\leq \phi \leq 2\pi$.
If we take the momentum of the external leg as a time-like vector and consider in the center-of-mass frame, we can set the momentum variables as follows:
\begin{align}
P(h_1)_{\mu}&=(\cos \tau_1,\sin \tau_1, 0, 0), \label{eq:ph1} \\
P(g)_{\mu}&=(x^{1/2}\cos \tau,x^{1/2}\sin \tau, (x-1)^{1/2}\cos \phi, (x-1)^{1/2}\sin \phi), \label{eq:pg}
\end{align}
where $x=\cosh^2 \rho$. Considering the condition (\ref{eq:idfphi}), it is enough to take the range of $\tau_1$ as $0\leq \tau_1 \leq \pi/2$ for the positive energy external leg as in Figure \ref{fig:identify}. 

Using these parameterizations, we obtain
\begin{align}
P^2(h_1^{-1}g)&=x\cos^2(\tau-\tau_1)-1, \\
\int dg&=\int_0^{2\pi}d\phi \int_0^{2\pi}d\tau \int _1^{\infty}\frac{dx}{2}.
\end{align}
Thus, the amplitude (\ref{eq:1loopamp}) becomes
\begin{align}
&i\Gamma_p^{(2)}=\lambda^2 \delta(h_1h_2)\int_0^{2\pi}d\phi \int_0^{2\pi}d\tau \int _1^{\infty}\frac{dx}{2}\frac{1}{x\cos^2 \tau -\cos^2 m}\frac{1}{x\cos^2 (\tau-\tau_1) -\cos^2 m} \notag \\
&=\lambda^2 \pi \delta(h_1h_2) \int_{0}^{2\pi}\frac{d\tau}{\cos^2\tau \cos^2(\tau-\tau_1)}\int_1^{\infty}\frac{dx}{(x-\cos^2m/\cos^2 \tau)(x-\cos^2m/\cos^2 (\tau-\tau_1))},
\end{align}
where we set $M=\sin m ~(0\leq m \leq \pi/2)$. For convenience, we integrate over $x$ from $1$ to $\Lambda$ and take the limit $\Lambda \to \infty$ later.\footnote{This is necessary for the $\tau$-integration to be carried out in a well-defined manner in the following.} Using the integral formula
\begin{equation}
\int_1^{\Lambda}\frac{dx}{(x-a)(x-b)}=\frac{1}{a-b}\ln\bigg(\frac{(1-b)(\Lambda-a)}{(1-a)(\Lambda-b)}\bigg),
\end{equation}
we find
\begin{align}
i\Gamma_p^{(2)}&=\lambda^2 \pi \delta (h_1h_2)\frac{1}{\cos^2m}\frac{1}{\sin \tau_1}\int_0^{2\pi} \frac{d\tau}{\sin 2\tau} \notag \\
&\cdot \ln\bigg[\frac{\sin(\tau -\frac{\tau_1}{2}+m)\sin(\tau-\frac{\tau_1}{2}-m)\sin(\tau+\tau_1/2+m_{\Lambda})\sin(\tau+\tau_1/2-m_{\Lambda})}{\sin(\tau +\frac{\tau_1}{2}+m)\sin(\tau+\frac{\tau_1}{2}-m)\sin(\tau-\tau_1/2+m_{\Lambda})\sin(\tau-\tau_1/2-m_{\Lambda})} \bigg],
\end{align}
where we have defined $\cos m_{\Lambda}=\cos m/\sqrt{\Lambda}$ and shifted $\tau$ to $\tau+\tau_1/2$.

Then, we consider the $\tau$-integral,
\begin{align}
I(\tau_1)&=\int_0^{2\pi} \frac{d\tau}{\sin 2\tau} \notag \\
&\cdot \ln\bigg[\frac{\sin(\tau -\frac{\tau_1}{2}+m)\sin(\tau-\frac{\tau_1}{2}-m)\sin(\tau+\tau_1/2+m_{\Lambda})\sin(\tau+\tau_1/2-m_{\Lambda})}{\sin(\tau +\frac{\tau_1}{2}+m)\sin(\tau+\frac{\tau_1}{2}-m)\sin(\tau-\tau_1/2+m_{\Lambda})\sin(\tau-\tau_1/2-m_{\Lambda})} \bigg].
\end{align}
Differentiating $I(\tau_1)$ with respect to $\tau_1$, we obtain
\begin{align}
I'(\tau_1)=\int_0^{2\pi} \frac{d\tau}{2\sin 2\tau}&\bigg[\frac{\cos(m-\tau-\frac{\tau_1}{2})}{\sin(m-\tau-\frac{\tau_1}{2})}-\frac{\cos(m_{\Lambda}-\tau-\frac{\tau_1}{2})}{\sin(m_{\Lambda}-\tau-\frac{\tau_1}{2})} \notag \\
&-\frac{\cos(m+\tau-\frac{\tau_1}{2})}{\sin(m+\tau-\frac{\tau_1}{2})}+\frac{\cos(m_{\Lambda}+\tau-\frac{\tau_1}{2})}{\sin(m_{\Lambda}+\tau-\frac{\tau_1}{2})} \notag \\
&+\frac{\cos(m-\tau+\frac{\tau_1}{2})}{\sin(m-\tau+\frac{\tau_1}{2})}-\frac{\cos(m_{\Lambda}-\tau+\frac{\tau_1}{2})}{\sin(m_{\Lambda}-\tau+\frac{\tau_1}{2})} \notag \\
&-\frac{\cos(m+\tau+\frac{\tau_1}{2})}{\sin(m+\tau+\frac{\tau_1}{2})}+\frac{\cos(m_{\Lambda}+\tau+\frac{\tau_1}{2})}{\sin(m_{\Lambda}+\tau+\frac{\tau_1}{2})} \bigg].
\end{align}
Replacing $\tau$ to $w\equiv e^{2i\tau}$, it becomes
\begin{align}
I'(\tau_1)=i \oint \frac{dw}{w^2-1}\bigg[&-\frac{w+\alpha_{-}}{w-\alpha_{-}}-\frac{w+\alpha_{-}^{-1}}{w-\alpha_{-}^{-1}}-\frac{w+\alpha_{+}}{w-\alpha_{+}}-\frac{w+\alpha_{+}^{-1}}{w-\alpha_{+}^{-1}} \notag \\
&+\frac{w+\beta_{-}}{w-\beta_{-}}+\frac{w+\beta_{-}^{-1}}{w-\beta_{-}^{-1}}+\frac{w+\beta_{+}}{w-\beta_{+}}+\frac{w+\beta_{+}^{-1}}{w-\beta_{+}^{-1}}\bigg] , \label{eq:I'integral}
\end{align}
where $\alpha_{\pm}\equiv e^{2i(m\pm\tau_1/2)},~\beta_{\pm} \equiv e^{2i(m_{\Lambda}\pm\tau_1/2)}$. Taking the $-i\epsilon$-prescription, $m,~m_{\Lambda}$ are shifted to $m-i\epsilon,~m_{\Lambda}-i\epsilon$, respectively. Thus, the poles which contribute to the contour integral are the only $w=\alpha_{-}^{-1}, \alpha_{+}^{-1},\beta_{-}^{-1}, \beta_{+}^{-1}$. Carrying out the contour integral, we obtain
\begin{equation}
I'(\tau_1)=-2\pi i \bigg(\frac{1}{\sin (\tau_1-2m)}-\frac{1}{\sin (\tau_1+2m)}- \frac{1}{\sin (\tau_1-2m_{\Lambda})}+\frac{1}{\sin (\tau_1+2m_{\Lambda})}\bigg).\label{eq:I'result}
\end{equation}
Taking the limit $\Lambda\to 0$, the last two terms in (\ref{eq:I'result}) are canceled because $m_{\Lambda}$ goes to $\pi/2$. Integrating $I'(\tau_1)$ over $\tau_1$ and using $I(0)=0$, we obtain
\begin{equation}
I(\tau_1)=-2\pi i \ln \bigg(\frac{\tan (m-\frac{\tau_1}{2})}{\tan (m+\frac{\tau_1}{2})}\bigg).
\end{equation}
Thus, the planar amplitude is 
\begin{equation}
i\Gamma_p^{(2)}=-i\frac{2\pi^2\lambda^2}{\cos^2m}\delta(h_1h_2)\frac{1}{\sin \tau_1}\ln \bigg(\frac{\tan (m-\frac{\tau_1}{2})}{\tan (m+\frac{\tau_1}{2})}\bigg). \label{eq:planarresult}
\end{equation}

\begin{figure}
\begin{center}
\includegraphics[scale=0.7]{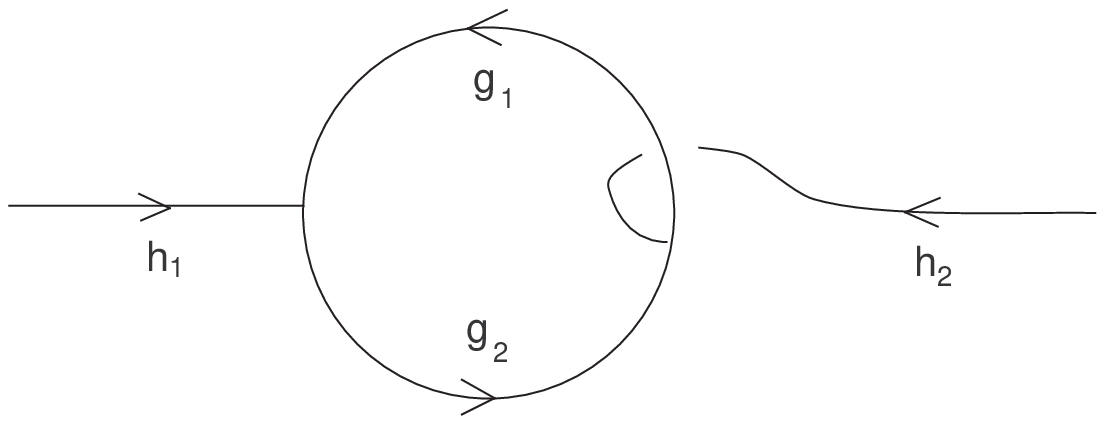}
\end{center}
\caption{The one-loop self-energy nonplanar diagram.}
\label{fig:nonplanar1-loop}
\end{figure}

We can also obtain the amplitude of the one-loop self-energy nonplanar diagram in the noncommutative $\phi^3$ theory, which is depicted as in Figure \ref{fig:nonplanar1-loop}.
Using the Feynman rules (\ref{eq:feynman rule}), the nonplanar amplitude is given by
\begin{align}
i\Gamma_{np}^{(2)}&=(-i\lambda)^2\int dg_1dg_2\frac{-i}{P^2(g_1)+M^2}\frac{-i}{P(g_2)^2+M^2}\delta(g_2^{-1}h_1g_1)\delta(g_2h_2g_1^{-1}) \notag \\
&=\lambda^2\int dg_1\frac{1}{P^2(g_1)+M^2}\frac{1}{P(g_2)^2+M^2}\delta(h_1g_1h_2g_1^{-1}). \label{eq:nonplanar}
\end{align}
The result is \cite{Imai:2000kq}
\begin{align}
i\Gamma_{np}^{(2)}&=\frac{\lambda^2\pi}{\sqrt{1-M^2}}\theta(-p_0p_0')\delta(p_{-1}-p_{-1}') \notag \\
&\cdot \frac{1}{\sqrt{(p+p')^2/4-M^2p^2}}\cdot \frac{2(1-M^2)p^2-(p+p')^2/4}{-((p+p')^2/4)^2-4(1-M^2)((p+p')^2/4-M^2p^2)}, \label{eq:nonplanar result}
\end{align}
where $p_i=P(h_1)_i,~p_i'=P(h_2)_i$.

\begin{figure}
\begin{center}
\includegraphics[scale=0.6]{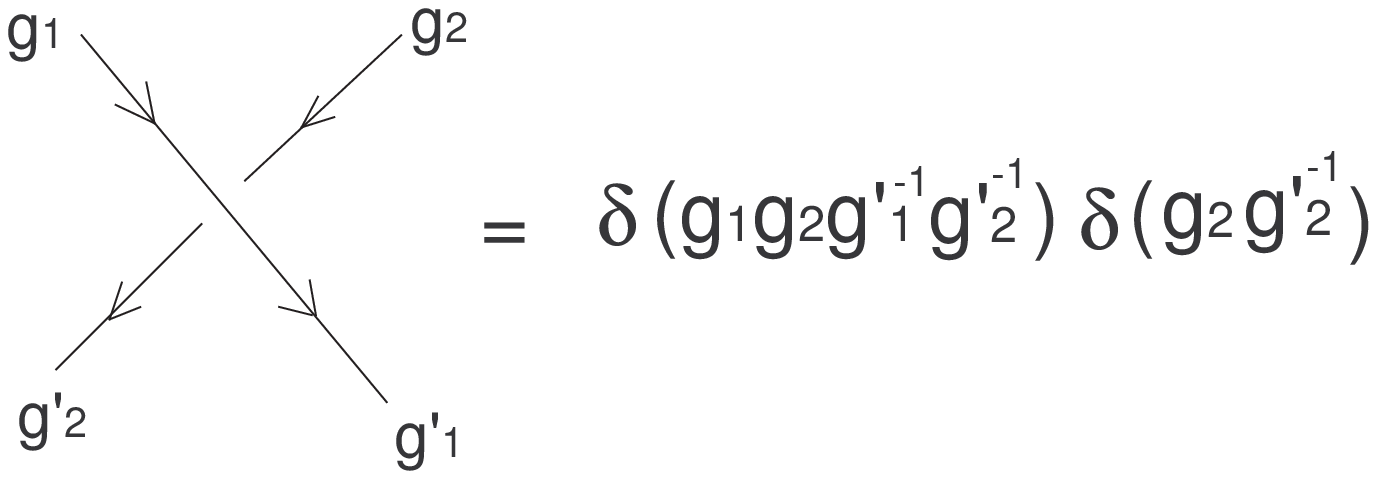}
\end{center}
\caption{The braiding rule.}
\label{fig:braidfeynmanrule}
\end{figure}

From the above expression, we find that the external momenta are not conserved. In general, we can see that nonplanar diagrams in noncommutative field theory in the Lie-algebraic noncommutative spacetime do not possess the external momentum conservation law. But as we described in the introduction, in order to possess a Hopf algebraic symmetry in a field theory at quantum level, we have to include a nontrivial statistics, which is called braiding \cite{Sasai:2007me}. In the case of the noncommutative scalar field theory in the Lie-algebraic noncommutative spacetime, the braiding is given by \cite{Freidel:2005bb,Sasai:2007me}
\begin{equation}
\psi(\ftilde_1(g_1)\otimes \ftilde_2(g_2))=\ftilde_2(g_2)\otimes \ftilde_1(g_2^{-1}g_1g_2), \label{eq:braid}
\end{equation}
where $\psi$ means the exchange of two fields. Thus, we should include the additional Feynman rule as in Figure \ref{fig:braidfeynmanrule}.

Considering the braiding rule, the nonplanar amplitude (\ref{eq:nonplanar}) becomes
\begin{align}
i\Gamma_{np}^{(2)}&=(-i\lambda)^2\int dg_1dg'_1dg_2\frac{-i}{P^2(g_1)+M^2}\frac{-i}{P(g_2)^2+M^2}\delta(g_2^{-1}h_1g_1)\delta(g_2h_2g_1^{'-1})\delta(g_1^{-1}h_2g_1'h_2^{-1}) \notag \\
&=(-i\lambda)^2\delta(h_1h_2)\int dg_1\frac{-i}{P^2(g_1)+M^2}\frac{-i}{P(h_1g_1)^2+M^2}. \notag
\end{align}
Since this is the same as the planar amplitude (\ref{eq:planarresult}), the momentum conservation is restored.

It is worth mentioning that even if fields possess a nontrivial braiding, we can formulate correlation functions by using the braided path integral \cite{Oeckl:1999zu,Sasai:2007me}. The naively derived Feynman rules (\ref{eq:feynman rule}) are justified in the context of the braided quantum field theory.

\subsection{The Cutkosky rule of the one-loop self-energy diagram} \label{subsec:unitarity}
We check whether the Cutkosky rule \cite{Cutkosky:1960sp,Peskin:1995ev}, which gives the unitary relation of S-matrix in conventional field theories, is satisfied in the noncommutative field theory in the Lie algebraic noncommutative spacetime at the one-loop self-energy diagram.

\begin{figure}
\begin{center}
\includegraphics[scale=0.5]{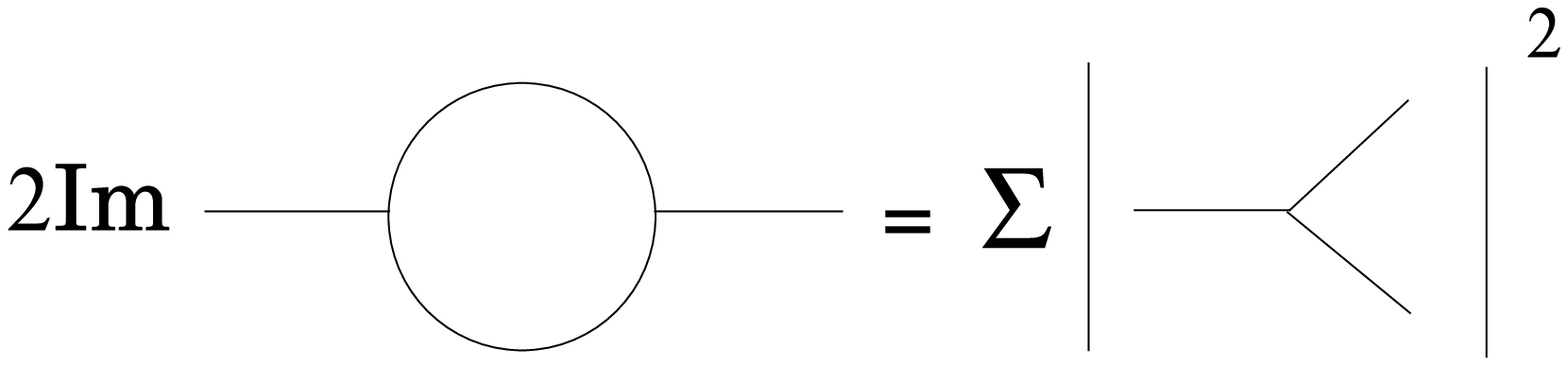}
\end{center}
\caption{The Cutkosky rule of the one-loop self-energy diagram in $\phi^3$ theory.}
\label{fig:cutkoski}
\end{figure}

The Cutkosky rule is given by
\begin{equation}
2\mathrm{Im} \Gamma_{ab} =\sum_n\Gamma_{an}\Gamma_{nb}^{\ast}, \label{eq:cutkoski}
\end{equation}
where $\Gamma_{ab}$ is the transition matrix element between states $a$ and $b$, and 
the summation is over all the ways to cut through the diagram such that
the cut propagators can simultaneously be put on shell. 
When we check the unitarity, we impose the on-shell conditions on the external legs,
where the on-shell conditions restrict the energies to reside on the bold line 
in Figure \ref{fig:identify}. In $\phi^3$ theory, the Cutkosky rule of the one-loop self-energy diagram is given by Figure \ref{fig:cutkoski}.

\begin{figure}
\begin{center}
\includegraphics[scale=0.7]{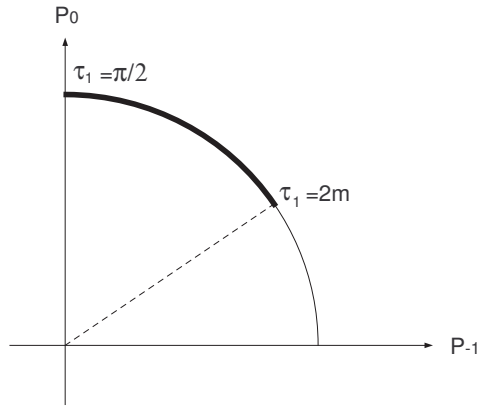}
\end{center}
\caption{The regions of the branch cuts in (\ref{eq:branch}) when $0\leq m\leq \frac{\pi}{4}$.}
\label{fig:branchcuts}
\end{figure}

As we have seen in the previous section, the one-loop nonplanar self-energy diagram becomes the same contribution as the planar diagram if we include the braiding (\ref{eq:braid}). Thus, we only check the Cutkosky rule of the planar diagram. The imaginary part of the planar amplitude (\ref{eq:planarresult}) is given by
\begin{equation}
2\mathrm{Im} \Gamma_p^{(2)}=i\frac{2\pi^2\lambda^2}{\cos^2 m}\delta(h_1h_2)\frac{1}{\sin \tau_1}\bigg(\ln \bigg(\frac{\tan (m-i\epsilon-\frac{\tau_1}{2})}{\tan (m-i\epsilon+\frac{\tau_1}{2})}\bigg)-\ln \bigg(\frac{\tan (m+i\epsilon-\frac{\tau_1}{2})}{\tan (m+i\epsilon+\frac{\tau_1}{2})}\bigg)\bigg). \label{eq:branch}
\end{equation}
This expression has branch cuts in the following regions:
\begin{align}
2m \leq \tau_1 \leq \frac{\pi}{2},~~~&\text{for}~0\leq m\leq \frac{\pi}{4}, \label{eq:imregion1} \\
\pi-2m\leq \tau_1\leq \frac{\pi}{2},~~~&\text{for}~\frac{\pi}{4}\leq m\leq \frac{\pi}{2}, \label{eq:imregion2} 
\end{align}
because the arguments of the logarithm become negative. Figure \ref{fig:branchcuts} shows the region of the branch cut in (\ref{eq:branch}) when $0\leq m\leq \pi/4$. 

\begin{figure}
\begin{center}
\includegraphics[scale=0.8]{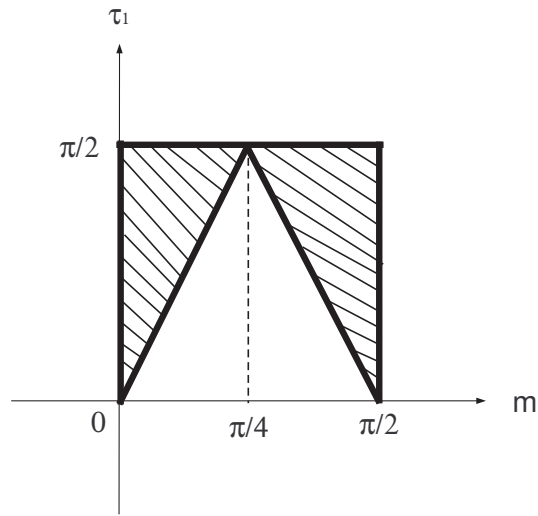}
\end{center}
\caption{The shadow areas show the regions in which (\ref{eq:branch}) does not vanish.}
\label{fig:imagrange}
\end{figure}

Evaluating the discontinuity of the Riemann surface, we obtain
\begin{equation}
2\mathrm{Im} \Gamma_p^{(2)}=\frac{4\pi^3\lambda^2}{\cos^2 m}\delta(h_1h_2)\frac{1}{\sin \tau_1}, \label{eq:imaginarypart}
\end{equation}
if $\tau_1$ is in the region given by (\ref{eq:imregion1}) or (\ref{eq:imregion2}). Otherwise, the imaginary part of the amplitude vanishes. Figure \ref{fig:imagrange} shows the regions in which (\ref{eq:branch}) does not vanish.

\begin{figure}
\begin{center}
\includegraphics[scale=0.6]{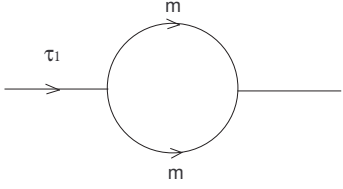}
\end{center}
\caption{The physical process for (\ref{eq:imregion1}).}
\label{fig:event1}
\end{figure}

\begin{figure}
\begin{center}
\includegraphics[scale=0.6]{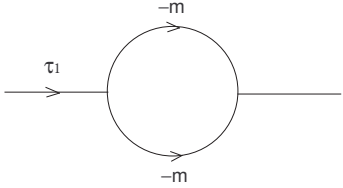}
\end{center}
\caption{The unphysical process for (\ref{eq:imregion2}).}
\label{fig:event2}
\end{figure}

We can give the physical interpretation of the result. If $m$ is less than $\pi/4$, the physical process given by Figure \ref{fig:event1} will contribute to the imaginary part of the amplitude, and the threshold for $\tau_1$ is $2m$, corresponding to the region (\ref{eq:imregion1}). On the other hand, if $m$ is larger than $\pi/4$, the unphysical process given by Figure \ref{fig:event2} will contribute to the imaginary part of the amplitude, because the threshold value of two negative masses $\pi-2m\sim -2m$ under the identification $g\sim -g$ is in the positive energy region. This corresponds to (\ref{eq:imregion2}).

To obtain the right hand side of (\ref{eq:cutkoski}), we replace the propagators in (\ref{eq:1loopamp}) by
\begin{equation}
\frac{1}{P^2(g)+M^2}\to 2\pi i\delta(P^2(g)+M^2),
\end{equation}
where we have to take only the positive energy poles for the direction of time. Since $g$ is identified with $-g$, the positive energy conditions are given by
\begin{align}
P_0(g)\geq 0~~&\text{for}~~P_{-1}(g)\geq 0,~~~~~~~~~~~~~~~~P_0(g)\leq 0~~\text{for}~~P_{-1}(g)\leq 0 \label{eq:positiveg} \\
P_0(g^{-1}h_1)\geq 0~~&\text{for}~~P_{-1}(g^{-1}h_1)\geq 0,~~~~P_0(g^{-1}h_1)\leq 0~~\text{for}~~P_{-1}(g^{-1}h_1)\leq 0. \label{eq:positivegh-1}
\end{align}
Using the parameterizations (\ref{eq:ph1}) and (\ref{eq:pg}), the positive energy conditions are represented as
\begin{align}
R: 0\leq \tau\leq \tau_1,~~\pi\leq \tau\leq \pi+\tau_1.
\end{align}

Then, the right hand side of (\ref{eq:cutkoski}) becomes
\begin{align}
&\sum |\Gamma |^2 \notag \\
&=4\pi^2 \lambda^2\delta(h_1h_2)\int_{R}d\tau \int_0^{2\pi}d\phi \int_1^{\infty}\frac{dx}{2}\delta(x\cos^2\tau-\cos^2m)\delta(x\cos^2(\tau-\tau_1)-\cos^2m) \notag \\
&=8\pi^3 \lambda^2\delta(h_1h_2)\int_1^{\infty}\frac{dx}{2x^2}\int_{R}d\tau \delta(\cos^2\tau-\cos^2m_x)\delta(\cos^2(\tau-\tau_1)-\cos^2m_x) \notag \\
&=4\pi^3 \lambda^2\delta(h_1h_2)\int_1^{\infty}\frac{dx}{x^2}\int_{R}d\tau \delta(\cos^2\tau-\cos^2m_x)\delta(\sin(m_x-\tau_1+\tau)\sin(m_x+\tau_1-\tau)), \label{eq:rhs of cutkoski1}
\end{align}
where $\cos m_x\equiv \cos m/\sqrt{x}$.
From the first delta-function in (\ref{eq:rhs of cutkoski1}), the possible values of $\tau$ are
\begin{equation}
\tau=m_x,~ \pi-m_x, ~\pi+m_x, ~2\pi-m_x.
\end{equation}

\begin{figure}
\begin{center}
\includegraphics[scale=0.65]{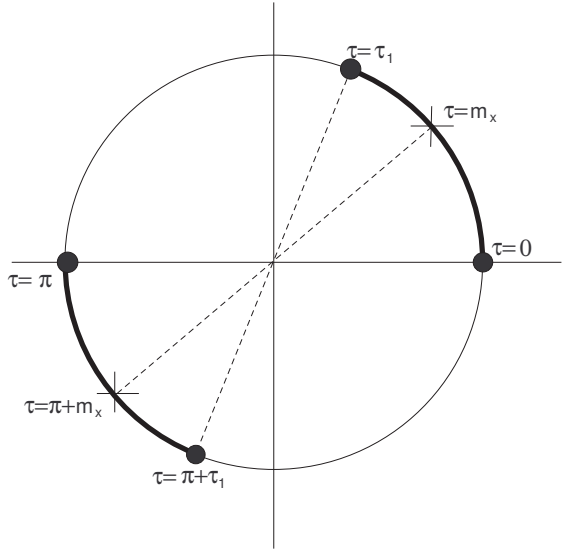}
\end{center}
\caption{The bold lines show the range of $R$. For (\ref{eq:rhs of cutkoski1}) to be non-zero, $m_x$ must be in the range of $R$.}
\label{fig:t1range}
\end{figure}

Since the range of $m$ is $0\leq m \leq \pi/2$, $m_x$ is in
\begin{equation}
m\leq m_x \leq \frac{\pi}{2}. \label{eq:rangemx1}
\end{equation}
Thus, $\tau=m_x$ and $\tau=\pi+m_x$ are in the range of $R$ if and only if $0\leq m\leq m_x\leq \tau_1$ as in Figure \ref{fig:t1range}. 

\begin{figure}
\begin{center}
\includegraphics[scale=0.7]{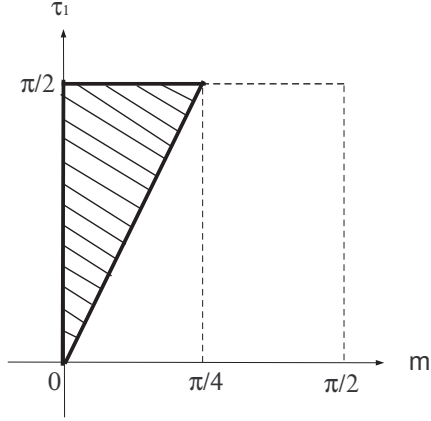}
\end{center}
\caption{The shadow areas show the regions in which the right hand side of (\ref{eq:cutkoski}) does not vanish.}
\label{fig:righthandrange}
\end{figure}

Taking these two values, (\ref{eq:rhs of cutkoski1}) becomes
\begin{align}
\sum |\Gamma |^2=4\pi^3\lambda^2 \frac{1}{\sin\tau_1}\delta(h_1h_2)\int_1^{\infty}\frac{dx}{x^2}\frac{1}{\cos m_x\sin m_x}\delta(\sin(2m_x-\tau_1)). \label{eq:seconddelta}
\end{align}
From the delta-function in (\ref{eq:seconddelta}), $m_x$ must satisfy
\begin{equation}
2m_x-\tau_1=n\pi,
\end{equation}
where $n\in \mathbb{Z}$. But from (\ref{eq:rangemx1}) and the range of $\tau_1$, the possible value of $n$ is $n=0$. Also, we find that $\tau_1$ is restricted to the following region:
\begin{equation}
2m\leq \tau_1 \leq \frac{\pi}{2}.
\end{equation}
Therefore, $m$ is restricted to the range of $0\leq m\leq \pi/4$. Integrating over $x$, we obtain
\begin{align}
\sum |\Gamma |^2=\frac{4\pi^3 \lambda^2}{\sin\tau_1\cos^2m}\delta(h_1h_2), \label{eq:gamma2}
\end{align}
which is the same as (\ref{eq:imaginarypart}). Thus, the right hand side of (\ref{eq:cutkoski}) is given by (\ref{eq:gamma2}) only if $2m\leq \tau_1 \leq \pi/2$ and $0\leq m\leq \pi/4$ as in Figure \ref{fig:righthandrange}. Otherwise, it is zero.

\begin{figure}
\begin{center}
\includegraphics[scale=0.7]{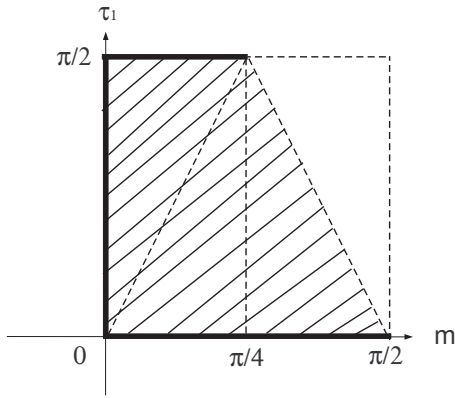}
\end{center}
\caption{The Cutkosky rule is satisfied in the shadow region.}
\label{fig:result}
\end{figure}

Comparing Figure \ref{fig:imagrange} and \ref{fig:righthandrange}, we find that the Cutkosky rule is satisfied  if 
\begin{align}
&0\leq \tau_1\leq \frac{\pi}{2}, ~~~\text{for}~~0\leq m \leq \frac{\pi}{4}, \notag \\
&0\leq \tau_1<\pi-2m, ~~~\text{for}~~\frac{\pi}{4}< m \leq \frac{\pi}{2}, \label{eq:result1}
\end{align}
and is violated if
\begin{align}
\pi-2m\leq \tau_1\leq \frac{\pi}{2}, ~~~\text{for}~~\frac{\pi}{4}< m \leq \frac{\pi}{2}, \label{eq:result2}
\end{align}
depicted as in Figure \ref{fig:result}.

\section{Summary and comment}
We have investigated the one-loop unitarity of the three dimensional braided noncommutative $\phi^3$ theory in the Lie algebraic noncommutative spacetime $[\xhat^i, \xhat^j]=2i\kappa \epsilon^{ijk}\xhat_k$ by examining the Cutkosky rule of the one-loop self-energy diagram. We did not have to evaluate the nonplanar amplitude because if we include the braiding, it has the same contribution as the planar one. Then, we have found that the Cutkosky rule is satisfied at the one-loop level when the mass $M$ is smaller than $1/\sqrt{2}\kappa$. This result is contrary to the fact that noncommutative field theories in the Moyal plane  violate the unitarity at the one-loop level when the time-like noncommutativity does not vanish irrespective of the values of mass.

However, the Cutkosky rule is found to be violated when the mass $M$ is larger than $1/\sqrt{2}\kappa$.  This enigmatic result comes from the fact that the virtual negative energy 
process depicted as in Figure \ref{fig:event2} occurs in the planar diagram. 
This mechanism of the violation of unitarity is different from that
in the Moyal-type noncommutative field theories with a non-zero time-like 
noncommutativity.

The above results, however, 
do not imply that the theory is unitary when the mass is smaller than
$1/\sqrt{2}\kappa$. 
Throughout this paper, we have only checked the Cutkosky rule of the one-loop self-energy amplitude. In more complicated amplitudes, the virtual negative energy processes occur more likely and the  
the Cutkosky rules will be broken for a smaller mass $M$, and 
the unitarity of the theory as a whole will be violated for any values of the mass. 
On the other hand, 
since this violation of the unitarity comes from the periodic property of the 
$SL(2,R)/Z_2$ group momentum space, 
the extension of the group momentum space to the universal 
covering group may drastically remedy the unitarity property of the theory.
This should be investigated in future works.

\section*{Acknowledgments}
Y.S. was supported in part by JSPS Research Fellowships for Young Scientists.
N.S. was supported in part by the Grant-in-Aid for Scientific Research No.~18340061 from the Ministry of Education, Science, Sports and Culture of Japan.

\appendix
\section{The proof of the formula (\ref{eq:pecom})} \label{sec:proofpecom}
Since the commutation relation between coordinates and momenta is written by only momenta, we can find
\begin{equation}
e^{-isk\cdot \xhat}\Phat_{\mu}e^{isk\cdot \xhat}=T_{\mu}{}^{\nu}(k;s)\Phat_{\nu}, \label{eq:com1}
\end{equation}
where $s$ is a real parameter. Differentiating both hands sides with respect to $s$, we obtain
\begin{equation}
-isk^i[\xhat_i,e^{-isk\cdot \xhat}\Phat_{\mu}e^{isk\cdot \xhat}]=\frac{d}{ds}T_{\mu}{}^{\nu}(k;s)\Phat_{\nu}. 
\end{equation}
Using (\ref{eq:com1}), the above equation becomes
\begin{equation}
\frac{d}{ds}T_{\mu}{}^{\nu}(k;s)\Phat_{\nu}=-isk^iT_{\mu}{}^{\nu}(k;s)[\xhat_i,\Phat_{\nu}]. \label{eq:deqofm}
\end{equation}
Using (\ref{eq:xhat}) and (\ref{eq:comjp}), the commutator between $\xhat_i$ and $\Phat_{\nu}$ becomes
\begin{equation}
[\xhat_i,\Phat_{\nu}]=i\kappa(-\eta_{-1,\nu}\Phat_i+\eta_{i\nu}\Phat_{-1}+\epsilon_{i}{}^{jk}\eta_{j\nu}\Phat_{k}).
\end{equation}
For convenience, we set $\kappa k^i\equiv \kbar^i$. We can write the equation (\ref{eq:deqofm}) as follows:
\begin{equation}
\frac{d}{ds}T_{\mu}{}^{\nu}(k;s)=T_{\mu}{}^{\rho}(k;s)K_{\rho}{}^{\nu}, \label{eq:difeqofm}
\end{equation}
where
\begin{equation}
K_{\rho}{}^{\nu}=
\begin{pmatrix}
0 & \kbar^0& \kbar^1& \kbar^2 \\
-\kbar^0 & 0 & -\kbar^2 & \kbar^1 \\
\kbar^1 & -\kbar^2 & 0 & -\kbar^0 \\
\kbar^2 &\kbar^1 & \kbar^0& 0 \\
\end{pmatrix}.
\end{equation}
The matrix $K$ is written by Pauli matrices as follows:
\begin{equation}
K=-\kbar^{1}\bar{\sigma}^{1}-\kbar^{2}\bar{\sigma}^{2}-\kbar^{3}\bar{\sigma}^{3}
\end{equation}
where 
\begin{align}
\bar{\sigma}^{1}&=-1\otimes \sigma^{1}, \notag \\
\bar{\sigma}^{2}&=\sigma^{2}\otimes \sigma^{2}, \notag \\
\bar{\sigma}^{3}&=-\sigma^{2}\otimes \sigma^{3}, \label{eq:pauli}
\end{align}
and $\kbar^3\equiv i\kbar^0$. $\bar{\sigma}^i$ follows the same relation as the Pauli matrices. Thus we can solve the equation (\ref{eq:difeqofm}). The solution is 
\begin{align}
(T)_{\mu}{}^{\nu}=(e^K)_{\mu}{}^{\nu}=\bigg(\cosh (\sqrt{\kbar^2})-\frac{\sinh (\sqrt{\kbar^2})}{\sqrt{\kbar^2}}\kbar^i\bar{\sigma}^i\bigg)_{\mu}{}^{\nu}
.
\end{align}
Using the expression (\ref{eq:pkrelation}), the matrix $M$ is represented by
\begin{equation}
T_{\mu\nu}=
\begin{pmatrix}
-P_{-1} & ~\Pbar_{0} & ~\Pbar_{1} & ~\Pbar_{2} \\
-\Pbar_{0} & ~-P_{-1} & ~-\Pbar_{2} & ~\Pbar_{1} \\
-\Pbar_{1} & ~\Pbar_{2} & ~P_{-1} & ~\Pbar_{0} \\
-\Pbar_{2} & ~-\Pbar_{1} & ~-\Pbar_{0} & ~P_{-1} \\
\end{pmatrix},
\end{equation}
where $\Pbar_{i}=\kappa P_i$.

\end{document}